\documentclass[a4paper,10pt,longbibliography, aps, prl,superscriptaddress]{revtex4-2}
\usepackage[utf8]{inputenc}

\usepackage{graphicx}
\usepackage{mathtools}
\usepackage{xcolor}
\usepackage{hyperref}
\usepackage{amsmath}
\usepackage{xfrac}
\usepackage{soul}
\usepackage{ulem}
\usepackage{float}

\begin{document}

	\title{A Traveling Wave Parametric Amplifier Isolator}
	\author{Arpit Ranadive} 
	\affiliation{Univ. Grenoble Alpes, CNRS, Grenoble INP, Institut N\'eel, 38000 Grenoble, France}
    \author{Bekim Fazliji} 
	\affiliation{Univ. Grenoble Alpes, CNRS, Grenoble INP, Institut N\'eel, 38000 Grenoble, France}
	\affiliation{Silent Waves, 38000 Grenoble, France}
    \author{Gwenael Le Gal} 
	\affiliation{Univ. Grenoble Alpes, CNRS, Grenoble INP, Institut N\'eel, 38000 Grenoble, France}
    \author{Giulio Cappelli} 
	\affiliation{Univ. Grenoble Alpes, CNRS, Grenoble INP, Institut N\'eel, 38000 Grenoble, France}
    \author{Guilliam Butseraen} 
    \thanks{Now at Silent Waves.}
	\affiliation{Univ. Grenoble Alpes, CNRS, Grenoble INP, Institut N\'eel, 38000 Grenoble, France}
    \author{Edgar Bonet} 
	\affiliation{Univ. Grenoble Alpes, CNRS, Grenoble INP, Institut N\'eel, 38000 Grenoble, France}
    \author{Eric Eyraud} 
	\affiliation{Univ. Grenoble Alpes, CNRS, Grenoble INP, Institut N\'eel, 38000 Grenoble, France}
    \author{Sina B\"{o}hling} 
    \affiliation{Institute for Theory of Condensed Matter and Institute for Quantum Materials and Technology, Karlsruhe Institute of Technology, 76131 Karlsruhe, Germany}  
    \author{Luca Planat}
    \affiliation{Silent Waves, 38000 Grenoble, France}
    \author{A. Metelmann} 
    \affiliation{Institute for Theory of Condensed Matter and Institute for Quantum Materials and Technology, Karlsruhe Institute of Technology, 76131 Karlsruhe, Germany} 
    \affiliation{Institut de Science et d’Ingénierie Supramoléculaires (ISIS, UMR7006), University of Strasbourg and CNRS}
    \author{Nicolas Roch} 
    \thanks{Corresponding author: nicolas.roch@neel.cnrs.fr}
	\affiliation{Univ. Grenoble Alpes, CNRS, Grenoble INP, Institut N\'eel, 38000 Grenoble, France}
    \affiliation{Silent Waves, 38000 Grenoble, France}
	
	\begin{abstract}

		Superconducting traveling-wave parametric amplifiers have emerged as highly promising devices for near-quantum-limited broadband amplification of microwave signals and are essential for high quantum-efficiency microwave readout lines. 
        Built-in isolation, as well as gain, would address their primary limitation: lack of true directionality due to potential backward travel of electromagnetic radiation to their input port. 
        Here, we demonstrate a Josephson-junction-based traveling-wave parametric amplifier isolator. 
        It utilizes third-order nonlinearity for amplification and second-order nonlinearity for frequency upconversion of backward propagating modes to provide reverse isolation. 
        These parametric processes, enhanced by a novel phase matching mechanism, exhibit gain of up to 20~dB and reverse isolation of up to 30~dB over a static 3~dB bandwidth greater than 500~MHz, while keeping near-quantum limited added noise. 
        This demonstration of a broadband truly directional amplifier ultimately paves the way towards broadband quantum-limited microwave amplification lines without bulky magnetic isolators and with inhibited back-action.

	\end{abstract}
	
	\maketitle

        Near-quantum-limited amplifiers~\cite{aumentado_superconducting_2020,esposito_perspective_2021} play a key role in millikelvin electronics and superconducting circuit Quantum Electrodynamics (cQED)~\cite{blais_circuit_2021}. Applications include microwave quantum optics experiments \cite{Eichler.2012i9, Nakamura.2013, fraudetDirectDetectionDownconverted2024}, superconducting \cite{stehlik_fast_2015,krantz_quantum_2019} and spin-qubits readout \cite{Schaal.2020, elhomsyBroadbandParametricAmplification2023}, measurement of nano-electromechanical systems \cite{teufel_sideband_2011}, electron spin resonance detection \cite{bienfait_reaching_2016}, astronomy instrumentation \cite{smith_low_2013, bockstiegel_development_2014}, or axionic dark matter detection \cite{jeong_search_2020, braineExtendedSearchInvisible2020, Grenet.2021, the_madmax_collaboration_simulating_2021,di_vora_search_2023}. A typical microwave measurement setup consists of several amplifiers, with the noise performance of the amplification chain predominantly determined by the noise and gain characteristics of the first amplifier \cite{caves_quantum_1982,clerk_introduction_2010}. To achieve the best noise performance at low temperatures in the microwave range, quantum-limited system noise (1 photon) and high gain (typically $>$ 20~dB) in the first amplification stage are required. 
        For this purpose, superconducting traveling-wave parametric amplifiers (TWPAs) have emerged among the most promising devices \cite{aumentado_superconducting_2020,esposito_perspective_2021}, offering excellent performances over a wide bandwidth (several GHz), with added noise close to the standard quantum limit (SQL) \cite{macklin_nearquantum-limited_2015,malnou_three-wave_2021,planat_photonic-crystal_2020,ranadive_kerr_2022,fadavi_roudsari_three-wave_2023}.\\
        
        \noindent Amplification in TWPAs is achieved through a wave mixing process resulting from the interaction between traveling modes in a nonlinear medium; standard implementations are based on either a high-kinetic inductance superconducting material (KTWPAs) or a Josephson-junction-based meta-material (JTWPAs). 
        This medium is stimulated by a strong pump field at frequency $\omega_\text{p}$, leading to the amplification of a weak signal field at frequency $\omega_\text{s}$ and the generation of an idler field at frequency $\omega_\text{i}$. 
        Media showing second-order nonlinearity give rise to a three-wave-mixing (3WM) process involving three photons, and media exhibiting third-order nonlinearity enable a four-wave-mixing (4WM) process involving four photons. 
        For either process, a long-standing challenge in achieving large amplification in long nonlinear transmission lines is maintaining phase matching between the coupled modes to obtain efficient energy exchange \cite{esposito_perspective_2021}.\\
        
        \noindent TWPAs are naturally directional devices once pumped, providing amplification solely to the waves co-propagating with the pump. 
        However, the backward-propagating waves are transmitted without attenuation. 
        Consequently, due to inevitable internal and external impedance mismatches, they can exhibit significant reverse gain and backward emissions within and outside their amplification bandwidth, thus potentially radiating toward the device under test (DUT).
        Therefore, it is crucial to use additional microwave components, known as isolators, to effectively protect the DUT from the inevitable radiation of the readout lines. 
        Such components have a significant footprint inside dilution cryostats, hindering scalability, and their dissipative nature spoils the readout quantum efficiency. 
        It is then highly desirable to get rid of these isolators, and an exciting solution would  be to have a device based on superconducting materials offering both amplifying and isolating functionalities, significantly reducing the footprint and dissipation.\\
        
        \noindent A general way to achieve this missing isolating property without applying large magnetic fields is to engineer dissipation dependent on the propagation direction. 
        Many theoretical and experimental implementations leveraging this idea have been proposed \cite{kamal_noiseless_2011,  abdo_directional_2013, metelmann_nonreciprocal_2015, sliwa_reconfigurable_2015, lecocq_nonreciprocal_2017, ranzani_wideband_2017, chapman_widely_2017,zhang_magnetic-free_2021,beck_wideband_2023, kwende_josephson_2023, ramosDirectionalJosephsonTravelingwave2023, naghilooBroadbandMicrowaveIsolation2021a}. 
        However, few of these studies predict \cite{Ranzani.2014, metelmann_nonreciprocal_2015, ramosDirectionalJosephsonTravelingwave2023} or show \cite{abdo_directional_2013, sliwa_reconfigurable_2015, lecocq_nonreciprocal_2017} gain and reverse isolation simultaneously, and only within narrow frequency bands of about one or two linewidths of the implemented resonant parametric amplifiers ($\sim$10~MHz).
        In contrast, TWPAs can intrinsically offer directionality for parametric processes, originating from the phase-matching requirement between the involved traveling waves.  \\

\begin{figure}[htb]
			\centering
			\includegraphics[width=.75\textwidth]{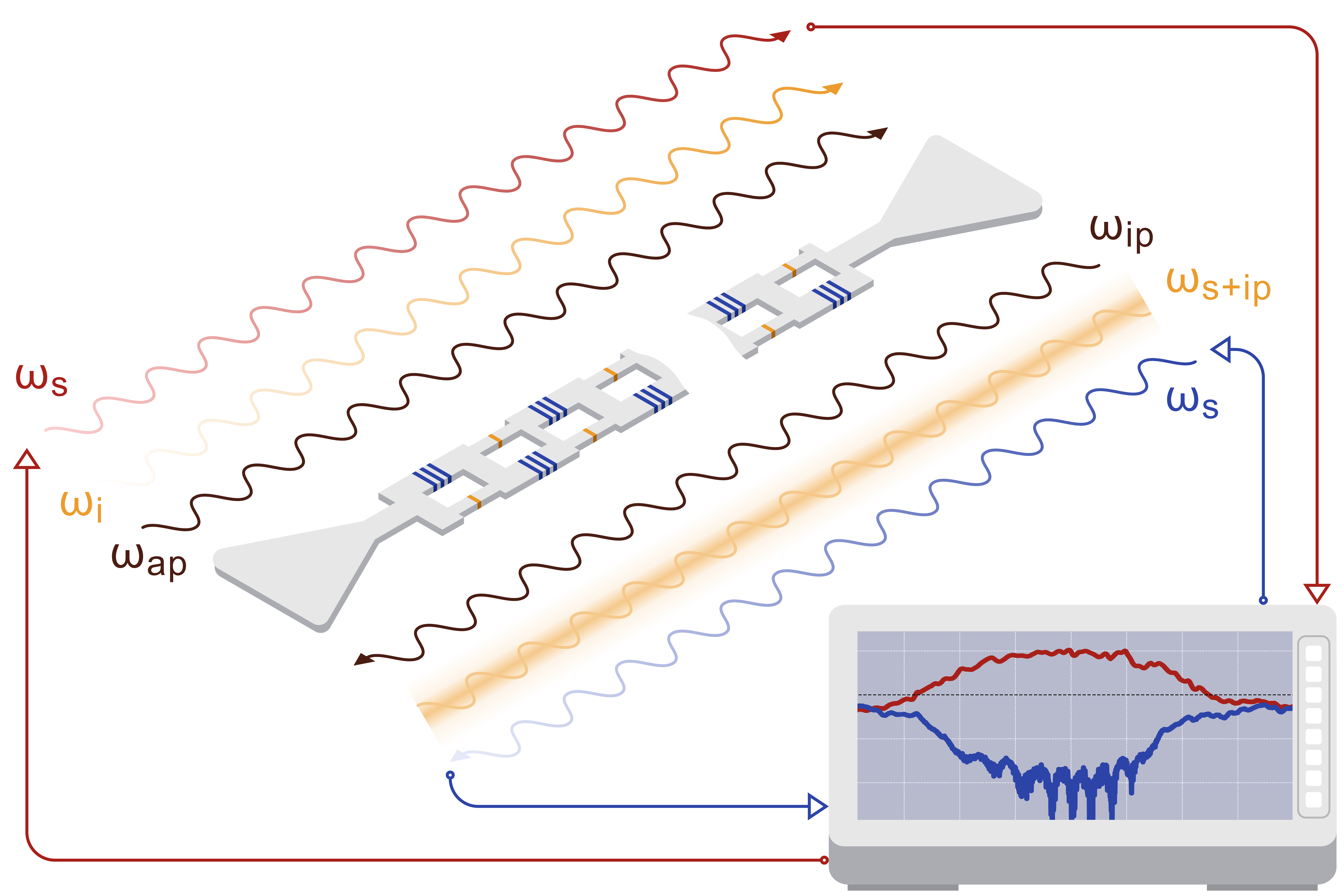}
            \caption{\textbf{Parametric amplification and isolation in the TWPAI.} 
            A high power right-moving pump tone at frequency $\omega_\text{ap}$ (brown) triggers four-wave-mixing parametric amplification of the right-moving signal tone at frequency $\omega_\text{s}$ as it propagates in the TWPAI. 
            The process is accompanied by the generation of an idler tone of frequency $\omega_\text{i}$ (yellow) for energy conservation. 
            Contrarily, a strong left-moving isolation pump tone at frequency $\omega_\text{ip}$ (brown) triggers frequency upconversion of the left-moving signal tone at frequency $\omega_\text{s}$ (identical to the amplified right-moving tone, blue). 
            The energy at $\omega_\text{s}$ is converted to a higher, out-of-band, frequency $\omega_\text{s+ip} = \omega_\text{s} + \omega_\text{ip}$ (yellow), leading to depletion, thus, reverse isolation at frequency $\omega_\text{s}$. 
            The propagation direction of the upconverted mode depends on the specifics of the process and the phase evolution of the involved modes; therefore, it is represented with a shaded wave.
            The middle part represents the chain of SNAILs, with three large Josephson junctions in one arm (blue) and a smaller one (orange) embedded in an Aluminum line (grey). 
            The SNAILs are inverted alternatively across the chain.
            At bottom right, a vector network analyzer displays forward (gain, red) and backward (isolation, blue) transmission of the device while double pumped.}
            \label{concepts}
		\end{figure}

    	\noindent The device presented here achieves amplification through third-order (or 4WM) nonlinearity. This parametric process ensures that efficient amplification only occurs for signals that travel in the same direction as the pump tone. 
        Consequently, there would be no backward or reverse amplification in an ideal scenario. 
        Additionally, we employ a backward-traveling second pump tone, which is used to implement a parametric frequency upconversion process based on the device's second-order nonlinearity (or 3WM)~\cite{ranzani_wideband_2017, naghilooBroadbandMicrowaveIsolation2021a}. 
        The unwanted electromagnetic wave ends up in a frequency band that is strongly detuned from the original one, thus achieving engineered depletion in the frequency domain.
        This frequency upconversion blocks any backward traveling radiation originating from either the measurement line or imperfect impedance matching from reaching the DUT, achieving reverse isolation.
        The required two nonlinearities for these processes are obtained simultaneously using superconducting nonlinear asymmetric inductive elements (SNAILs) \cite{frattini_3-wave_2017, frattini_optimizing_2018} as constitutive elements of the traveling-wave parametric amplifier isolator (TWPAI). 
        More specifically, the exact processes involved are depicted in Fig. \ref{concepts}: the right-moving waves at frequency $\omega_\text{s}$ are amplified as they propagate following a 4WM interaction with a right-moving pump at frequency $\omega_\text{ap}$ ($2\omega_\text{ap}=\omega_\text{s}+\omega_\text{i}$), while the left-moving waves at frequency $\omega_\text{s}$ are attenuated as the energy is upconverted to a higher frequency $\omega_\text{s+ip}=\omega_\text{s}+\omega_\text{ip}$ via a 3WM interaction with a left-moving pump tone at frequency $\omega_\text{ip}$. 
        The efficiency of these processes is ensured by utilizing a novel phase-matching mechanism based on the dynamics of wave propagation in the nonlinear medium with second-order nonlinearity.
        The TWPAI exhibits 20~dB directional gain and 30~dB reverse isolation with added noise near the standard quantum limit over a 500~MHz bandwidth.\\

	\section{Device description}
 
		The device discussed in this article is constituted of SNAIL superconducting loops with alternating orientation as depicted in Fig. \ref{concepts}, 
        Each SNAIL loop consists of three large Josephson junctions with a higher critical current, $I_0$, and one small Josephson junction with a lower critical current, $rI_\text{0}$ ($r < 1$), in either arm. 
		This metamaterial has been successfully adopted for four-wave-mixing amplification with reversed-Kerr phase matching \cite{ranadive_kerr_2022} and generation of broadband non-classical light \cite{esposito_observation_2022}.
		\\

		\noindent The asymmetry between the two arms of the superconducting loop enables the presence of both even and odd nonlinear terms in the Taylor expansion of the current-phase relation \cite{frattini_3-wave_2017},
        \begin{equation}
             I(\phi) \approx \frac{\Phi_\text{0}}{2\pi L} \phi - 3 \Phi_\text{0} \sqrt{\frac{R_\text{Q}}{\pi^3 Z^3}} g_\text{3} \phi^2 -  4 \Phi_\text{0} \frac{R_\text{Q}}{\pi^2 Z^2} g_\text{4} \phi^3 \, ,
             \label{current-phase}
			\end{equation}
		\noindent where $g_\text{3}$ and $g_\text{4}$ are the flux-tunable nonlinear coefficients for three-wave and four-wave-mixing, respectively, indicating the rates at which these interactions occur, $L$ represents the flux-tunable inductance per unit cell, $C_\text{g}$ the ground capacitance per unit cell, $\Phi_\text{0}$ the magnetic flux quantum, $R_\text{Q} = h/(4e^2)$ the resistance quantum, and $Z$ (defined as $\sqrt{L/C_\text{g}}$) is the characteristic impedance of the transmission line. 
        Explicit expressions are detailed in the methods section.
        The SNAIL naturally provides a much stronger second-order nonlinearity compared to the third-order nonlinearity \cite{frattini_3-wave_2017}. 
		To balance three-wave and four-wave-mixing processes, we use SNAIL elements with alternating magnetic flux polarity, i.e., they are oriented in a way such that magnetic flux biasing is reversed for adjacent SNAIL cells as shown in Fig. \ref{concepts}.
		The critical current ratio ($r$) serves as a pivotal design parameter for achieving an optimal balance between the two wave mixing processes within the device. 
		Moreover, applying a static magnetic field to the SNAILs allows for real-time adjustment of the wave mixing process strengths, thereby fine-tuning the device's operation to achieve the desired outcomes.
		\\

        \noindent The TWPAI contains 700 SNAIL cells, spanning 6~mm, with a critical current ratio of 0.07.
        It was fabricated on an intrinsic silicon substrate using the double-angle evaporation method with aluminum, followed by atomic layer deposition (ALD) of alumina dielectric and a copper top ground.
		The detailed fabrication process is discussed in \cite{planat_fabrication_2019,ranadive_kerr_2022,ranadive_nonlinear_2023}.
		\\

	\section{Gain optimization}
    
        \label{sec:gain_only}

        \begin{figure}[htb]
			\centering
			\includegraphics[width=0.95\textwidth]{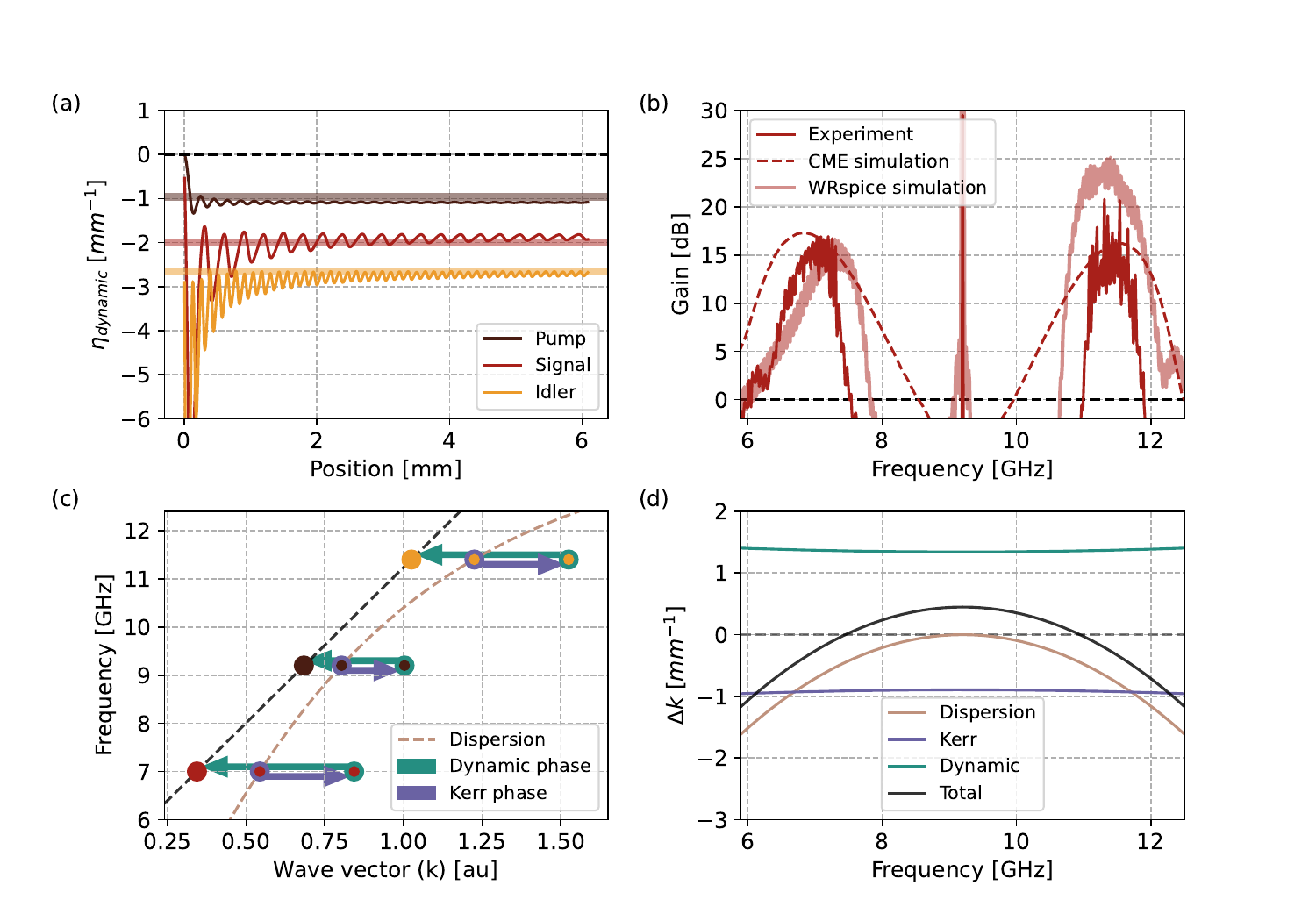}
            \caption{\textbf{Dynamic phase matching for amplification.}
            (a) The phase effect due to the presence of the second-order nonlinearity is shown, with solid curves representing solutions to the coupled mode equations and transparent-straight lines indicating dynamic coefficients ($\eta_\text{dynamic}$) computed using approximate expressions as detailed in the methods section. 
            Simulations and calculations are performed at Kerr-free flux to highlight the effect.
            (b) The experimental gain profile (solid curve) obtained by applying an external flux of 0.35~$\Phi_\text{0}$ and pumping the device ($\omega_\text{ap}$) at 9.2~GHz with $-78$~dBm power. 
            The gain simulation using coupled mode equations is shown as a dashed curve, and the WRspice simulation is shown as a transparent curve.
            (c)~Exaggerated dispersion relation to illustrate the phase matching mechanism.
            The three modes involved in four-wave-mixing amplification, signal (7~GHz), pump (9.2~GHz), and idler (11.4~GHz), are marked with the circles.
            The dashed peach line indicates dispersion in the medium, depicting curvature due to the plasma limit. 
            In the presence of Kerr nonlinearity, the dispersion is altered by cross-Kerr and self-Kerr effects (violet arrows), increasing the phase mismatch for the four-wave-mixing amplification process. 
            The dynamic phase effect, due to second-order nonlinearity, corrects this phase mismatch (green arrows), aligning the signal, pump, and idler modes on a straight line, achieving perfect phase matching.
            (d) The simulated phase mismatch (black) and constituting components, dispersion curvature (peach), Kerr effects (violet), and dynamic effects (green).
            }
            \label{fig:phase_matching}
		\end{figure}

		A significant difficulty for efficient wave-mixing in traveling-wave architecture is maintaining steady phase matching along the metamaterial. 
        For the four-wave-mixing amplification process, third-order or Kerr nonlinearity is utilized.
        This nonlinearity not only facilitates the essential coupling required for amplification but also introduces self-Kerr and cross-Kerr phase effects in the medium. 
        These phase effects alter the phase evolution of the traveling modes, contributing to phase mismatch. 
        Additionally, the curvature of the medium's dispersion, influenced by its plasma frequency, also affects phase mismatch. 
        Typically, these two sources of phase mismatch are engineered to counterbalance each other, thereby achieving phase matching for the amplification process \cite{macklinNearquantumlimitedJosephsonTravelingwave2015,planat_photonic-crystal_2020,ranadive_kerr_2022}.
        For the amplification process in the TWPAI, we solve the problem of phase matching by utilizing a novel mechanism involving dynamic phase effects due to the generation of higher harmonics and upconverted modes. 
        This approach does not require any additional complexity in the device design, maintaining the ease of fabrication.
        In the presence of a strong pump wave at frequency $\omega_\text{ap}$, three-wave-mixing within the medium leads to the generation of upconverted tones, 
        \begin{equation}
			A_\text{ap} \Leftrightarrow A_\text{2ap}, \hspace{1cm}
			A_\text{s} \Leftrightarrow A_\text{s+ap}, \hspace{1cm}
			A_\text{i} \Leftrightarrow A_\text{i+ap},
			\label{dynamic_pm1}
		\end{equation}
  
        \noindent where $A_\text{m}$ is amplitude of the wave  traveling at frequency $\omega_\text{m}$, with $m=ap,s,i$ etc. with frequencies signal $\omega_\text{s}$, idler $\omega_\text{i}$, amplification pump $\omega_\text{ap}$, $\omega_\text{2ap}=2\omega_\text{ap}$, $\omega_\text{s+ap}=\omega_\text{s}+\omega_\text{ap}$, and $\omega_\text{i+ap}=\omega_\text{i}+\omega_\text{ap}$.
        The phase mismatch between the modes limits the effective coupling, leading to a minimal energy exchange between them along the device.
        Although these conversion processes do not significantly deplete the modes, they impart an extra phase shift, influencing the phase evolution within the nonlinear medium. 
        A similar mechanism has already been discussed in the case of resonant structures made of SNAIL arrays and can be understood as a renormalisation of the Kerr-effect by three-wave-mixing processes~\cite{frattini_optimizing_2018}.
        The effective phase mismatch in presence of both, three-wave-mixing and four-wave-mixing nonlinearities can be expressed as a combination of three components,

		\begin{equation}
			\Delta k = \Delta k_{\text{dispersion}} + \Delta k_{\text{Kerr}} + \Delta k_{\text{dynamic}},
            \label{delta_k}
		\end{equation}

		\noindent where $\Delta k_\text{dispersion}$ and $\Delta k_\text{Kerr}$ are the usual phase mismatches coming from the curvature of the dispersion and the Kerr nonlinearity respectively and $\Delta k_\text{dynamic} = 2 \eta_\text{ap,dyn} - \eta_\text{s,dyn} - \eta_\text{i,dyn}$, with  $\eta_\text{ap/s/i,dyn}$ denoting the phase effects stemming from the dynamic processes listed in Eq. \ref{dynamic_pm1}. 
        The expressions of $\Delta k_{\text{dispersion}}$, $\Delta k_{\text{Kerr}}$, and $\eta_\text{ap/s/i,dyn}$ are detailed in the methods section.
        Figure \ref{dynamic_pm1} (a) depicts this dynamic effect obtained by numerically solving the coupled mode equations and their comparison with simplified expressions.
        In the TWPAI, dynamic phase mismatch $\Delta k_{\text{dynamic}}$ is optimized to cancel the phase mismatches arising from the dispersion curvature $\Delta k_{\text{dispersion}}$ and the Kerr effects $\Delta k_{\text{Kerr}}$, as depicted in Fig. \ref{fig:phase_matching} (c).
        This balance between phase mismatches ensures the phase-matched evolution of the waves involved in the four-wave-mixing amplification interaction, depicted in Fig. \ref{fig:phase_matching} (d); leading to the large gain shown in Fig. \ref{fig:phase_matching} (b).
        This amplification process can be qualitatively described using coupled mode equations detailed in the methods.
        However, to capture the finer details more accurately, a comprehensive circuit simulator like WRspice \cite{whiteley_josephson_1991_2} is needed.

    \section{Isolation optimization}

        \begin{figure}[htb]
			\centering
			\includegraphics[width=.495\textwidth]{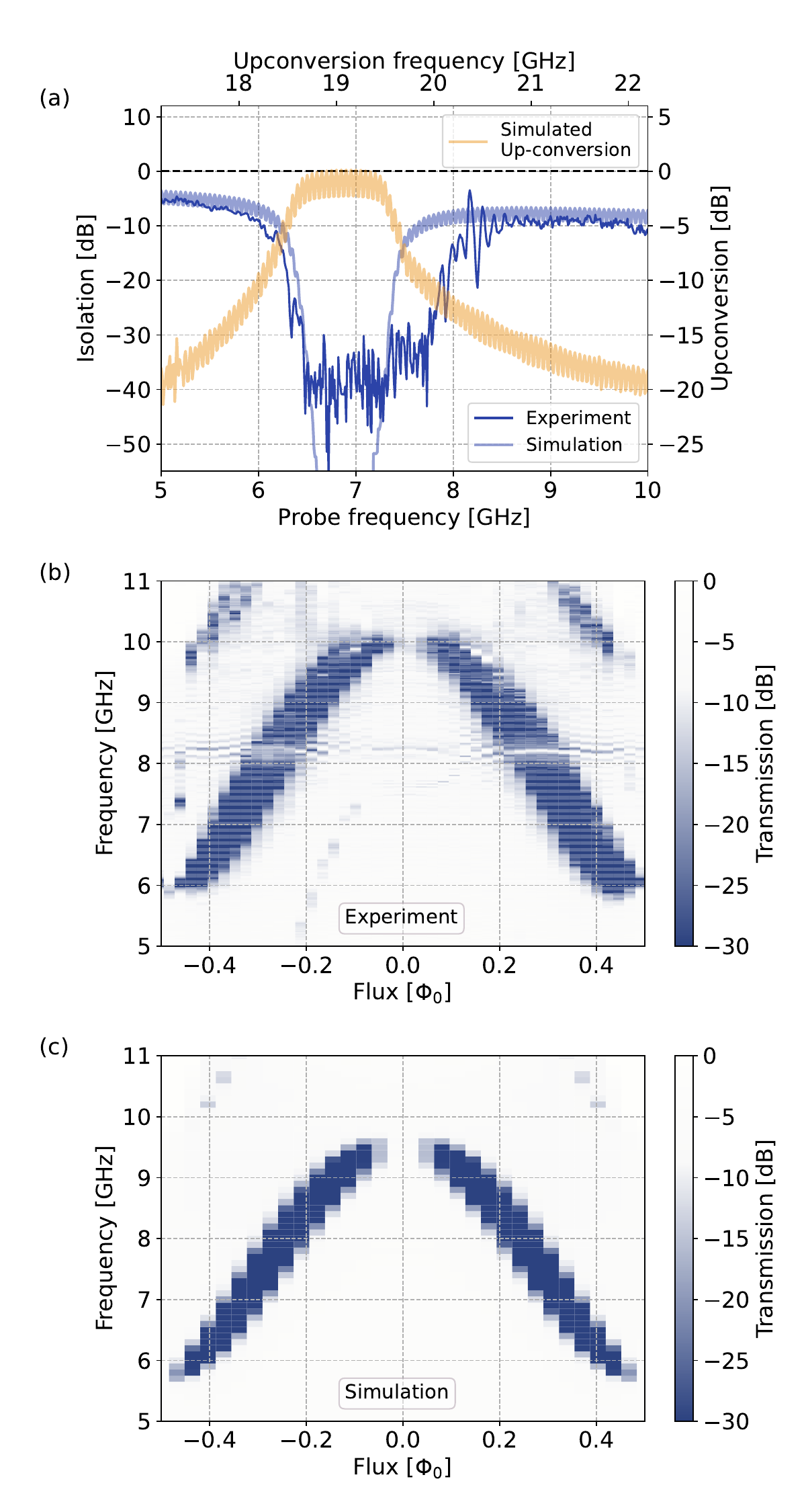}
            \caption{\textbf{Isolation utilizing three-wave-mixing upconversion.}
            (a) The experimental isolation profile (solid curve) obtained by applying an external flux of 0.35~$\Phi_\text{0}$ and pumping the device at $\omega_\text{ip}=12.2$~GHz with $-76$ dBm power. 
            The WRspice simulation for this isolation (due to upconversion) is shown as a transparent blue curve. 
            The total photon conversion to the frequency $\omega_\text{s}+\omega_\text{ip}$, referenced to the signal input, is indicated by the transparent yellow curve.
            (b) The measured isolation profile, resulting from upconversion, is shown as a function of the applied flux.
            (c) The corresponding WRspice simulations are depicted for comparison.}
            \label{fig:isolation}
		\end{figure}

        In addition to amplification, the device can also provide isolation through a three-wave-mixing upconversion process. 
        In this process, backward traveling waves at frequency $\omega_\text{s}$ combine with a pump tone at $\omega_\text{ip}$, resulting in the generation of waves at higher frequency $\omega_\text{s+ip} = \omega_\text{s} + \omega_\text{ip}$. 
        The dynamic phase-matching mechanism discussed in the case of amplification can also be helpful for the frequency upconversion process. 
        The modes involved in this case are,

        \begin{equation}
			A_\text{ip} \Leftrightarrow A_{2\text{ip}}, \hspace{1cm}
			A_\text{s} \Leftrightarrow A_\text{s+ip}, \hspace{1cm}
			A_\text{s+ip} \Leftrightarrow A_\text{s+2ip}.
			\label{dynamic_pm2}
		\end{equation}

        \noindent 
        The direction in which the upconverted mode propagates is dictated by the specific characteristics of the nonlinear interaction process and the phase evolution of the involved modes. 
        More precisely, it depends on the exact phase relationships between the isolation pump, signal, and upconverted tones and the medium's inherent properties, such as its dispersion and nonlinear phase effects. 
        Consequently, the upconverted mode can propagate either parallel or antiparallel to the isolation pump. Understanding the exact dependence of the propagation direction on these various processes is beyond the scope of this article.
        As some of the frequencies involved in the isolation process lie near the plasma frequency it was not feasible to model it using the simple coupled mode equations.
        However, a comprehensive circuit simulator like WRspice \cite{whiteley_josephson_1991_2} can simulate these parametric processes.
        Figure \ref{fig:isolation} (a) shows the experimentally obtained isolation (referenced to a lossless transmission line) at 0.35~$\Phi_0$ (blue), reaching -30~dB over a bandwidth of approximately 1.5 GHz. 
        The blue-shaded curve represents the WRspice simulation. 
        The yellow-shaded curve depicts the total photon flux (combining forward and backward propagation) at the upconverted frequency ($\omega_{\text{s+ip}}$) normalized to the probe photon flux. 
        At peak isolation, most of the input probe photons are upconverted, indicating that this upconversion process is primarily responsible for achieving isolation. 
        This observation confirms the mechanism behind the isolation, showing that the energy from the input probe is being effectively transferred to a higher frequency via the upconversion process. 
        A detailed study of this upconversion process will involve measurements at frequency bands outside the capacity of the available measurement setup (Fig. \ref{exp_setup}), and is beyond the scope of this article.
        Panels (b) and (c) compare the experimentally observed isolation with simulations at various fluxes. 
        The isolation vanishes at 0 and a half flux (i.e., where second-order nonlinearity is zero \cite{frattini_optimizing_2018}), as it relies on a three-wave-mixing process.

	\section{Gain with isolation, and noise characterization}

\begin{figure}[htb]
			\centering
			\includegraphics[width=.85\textwidth]{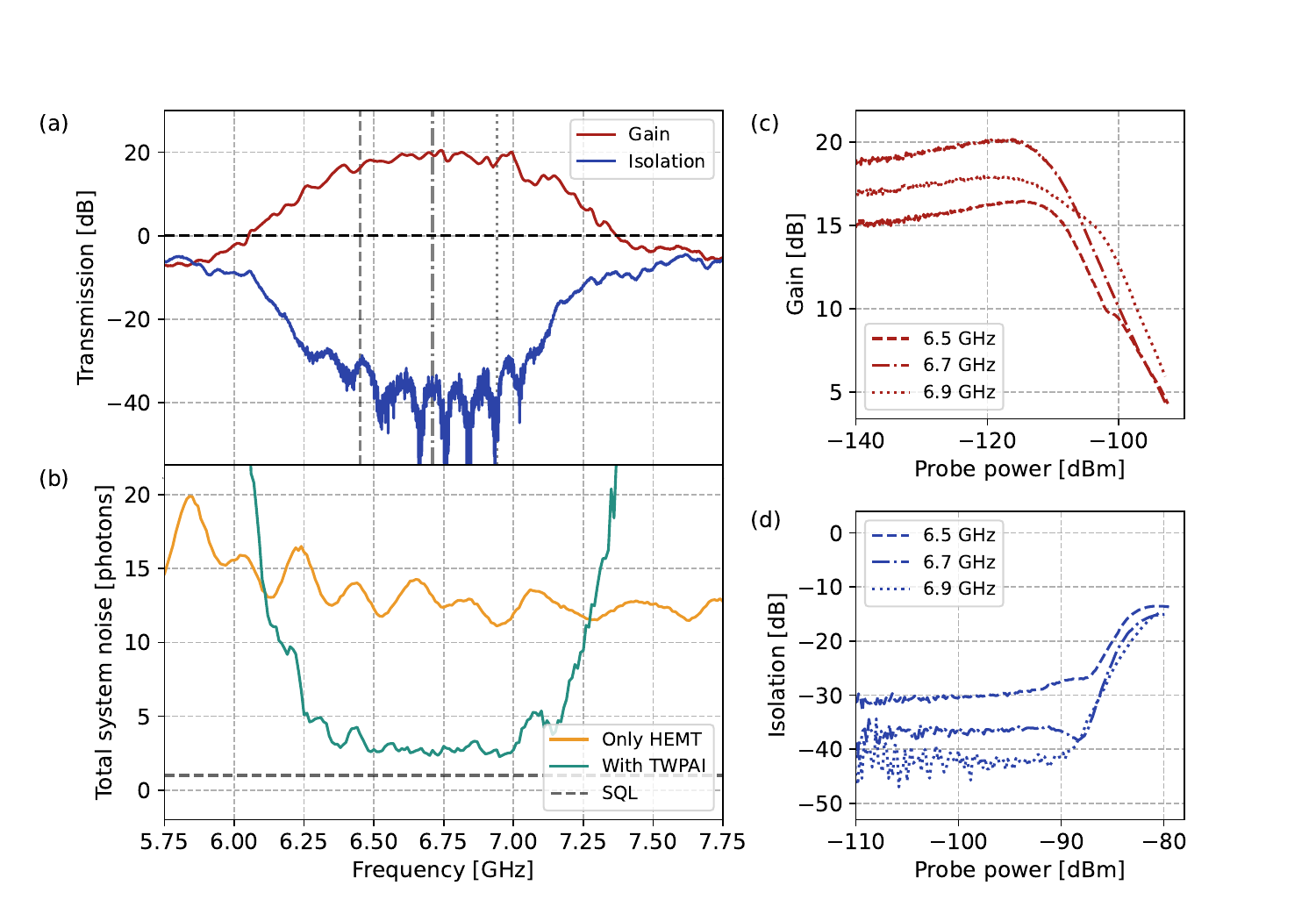}
            \caption{\textbf{Amplification with isolation.}
            (a) The measured forward transmission (gain, red) and backward transmission (isolation, blue) were obtained by applying an external flux of 0.19~$\Phi_\text{0}$ and double-pumping the device with a forward pump at 9.5~GHz with $-77$~dBm power and a backward pump at 13.85~GHz with $-79$~dBm power. 
            The device achieved a 20 dB gain over a bandwidth greater than 500~MHz and corresponding isolation greater than 30~dB, providing directionality exceeding 50~dB. 
            Both gain and isolation are referenced to a lossless transmission line.
            (b) The total system noise is shown for a double-pumped TWPAI (green) corresponding to the gain shown in panel (a), along with a reference (yellow) when the TWPAI is replaced with a lossless transmission line, and the standard quantum limit (SQL, grey dashed). 
            At peak gain, the total system noise is approximately three times the SQL, which is expected due to dielectric losses.
            (c) The evolution of gain with respect to the input signal power for signal frequencies highlighted in panel (a). 
            The gain saturation occurs between $-105$~dBm and $-110$~dBm of input signal power.
            (d) The evolution of isolation with respect to the input probe power for probe frequencies highlighted in panel (a).
            }
            \label{fig:gain}
		\end{figure}

		The TWPAI was experimentally characterized in a dilution refrigerator operating at 20~mK, employing a setup designed to facilitate measurements in both forward and backward directions \cite{ranzani_two-port_2013}, as detailed in the methods section (Fig. \ref{exp_setup}). 
		The device's performance, when flux biased at 0.19~$\Phi_\text{0}$, is illustrated in Fig. \ref{fig:gain}. 
		Panel (a) showcases the gain and isolation characteristics, measured as the difference in transmission when the device is pumped, referenced to a lossless equivalent transmission line.  
		These curves were generated through double-pumping the device: a forward amplification pump ($\omega_\text{ap}$) was applied at 9.5~GHz alongside a backward isolation pump ($\omega_\text{ip}$) at 13.8~GHz, as illustrated in Fig. \ref{concepts}.
		The forward transmission (depicted in red in Fig. \ref{fig:gain}) demonstrates an approximate 20~dB gain across a bandwidth larger than 500~MHz utilizing the four-wave-mixing amplification with the dynamic phase-matching introduced in the previous section. 
		The backward transmission (depicted in blue in Fig. \ref{fig:gain}) displays isolation up to 30~dB within the same band as the gain due to three-wave-mixing upconversion of backward traveling modes.
		The amplification curve depicted here corresponds to the lower lobe of the gain profile (Fig. \ref{fig:phase_matching} (b) shows the double-lobed shape of the gain profile).\\
  
		\noindent Josephson TWPAs exhibit high efficiency in amplification and require minimal pump power to operate effectively.
		Pump powers of approximately $-77$~dBm and $-79$~dBm were used for the amplification and isolation pumps, respectively, at the input/output of the TWPAI to achieve the performance depicted in Fig. \ref{fig:gain}.
		In the case of double pumping, the maximum power for each pump mode is inherently lower than that of a similar device operating in a single pump configuration. 
		Similar Josephson junction-based TWPAs are typically operated with a pump power of approximately $-75$~dBm \cite{planat_photonic-crystal_2020,ranadive_kerr_2022}. 
		This reduction in available pump power within a TWPAI naturally leads to decreased saturation power or 1~dB compression point of the gain.
		Saturation curves for the TWPAI are shown in Fig. \ref{fig:gain} (c) and in Fig. \ref{fig:gain} (d) for amplification and isolation, respectively. 
		The amplification process reaches 1~dB compression at an input signal power of approximately $-108$~dBm. 
		While this is lower than typically achieved in state-of-the-art Josephson-junction-based TWPAs, it remains sufficiently high to accommodate a significant potential for multiplexed measurement setups.
		Given the efficient nature of the upconversion process employed for isolation, the saturation power for isolation proves to be substantially higher. 
		It exceeds the necessary threshold to maintain truly directional amplification across a broad range of applied signal powers. \\

		\noindent A significant advantage of the methods for achieving directional gain utilized to realize the TWPAI is that they do not rely on any dispersion features within the device. 
		This minimizes gain ripples in amplification stemming from impedance mismatches near the dispersion features \cite{planat_photonic-crystal_2020,ranadive_kerr_2022}. 
		Furthermore, the phase matching in these devices is not contingent upon a fixed design parameter, thus allowing for in-situ tuning of the amplification band over gigahertz by adjusting the frequency of the amplification and isolation pumps (see Fig. \ref{tunability} in methods).
		Magnetic flux applied to the SNAILs can serve as an additional tuning parameter to optimize directional characteristics at various frequencies, resulting in even more extensive amplification band tunability. 
		However, a comprehensive description of flux tunability falls beyond the scope of the current discussion.\\

        \noindent The noise temperature and the gain of the amplification chain without the TWPAI were measured using a broadband thermal noise source with temperature tunability between 40~mK and 1~K (detailed in methods). 
		The noise performance of the amplification chain with TWPAI was calculated indirectly by measuring the noise emission of the amplification chain in the absence of any input in the amplification band and then normalizing it by the known TWPAI and system gain.
		The noise temperature of the system containing the TWPAI can be expressed as,

		\begin{equation}
			N_{\text{TWPAI}}(\omega) = \frac{P(\omega)}{G_\text{TWPAI} G_{\text{sys}} B_\text{w}},
		\end{equation}

		\noindent where $N_{\text{TWPAI}}(\omega)$ is the total noise of the measurement chain containing the TWPAI, $P(\omega)$ is the output power of the measurement chain with TWPAI input terminated with cold load, $G_{\text{TWPAI}}$ is the gain of the TWPAI including its insertion losses (referenced to lossless transmission), $G_{\text{sys}}$ is the gain of the measurement setup excluding TWPAI, and $B_\text{w}$ denotes measurement bandwidth.
		Figure \ref{fig:gain} (b) depicts the noise performance of the system containing double-pumped TWPAI (green), the noise performance of the system without TWPAI (yellow), dominated by the HEMT amplifier, and the standard quantum limit of the total system noise (dashed gray).
		At peak gain, the noise temperature of the system containing TWPAI is approximately three times the standard quantum limit. 
		This is comparable to amplifiers fabricated utilizing similar technology \cite{planat_photonic-crystal_2020,ranadive_kerr_2022}. 
		This extra added noise can be attributed to the presence of dissipation in the dielectric used for fabricating the capacitors in the device \cite{planat_fabrication_2019,ranadive_nonlinear_2023}.

	\section{Efficacy of the isolation}

		TWPAs inherently exhibit directional characteristics, i.e. they only amplify signals co-propagating with the pump tone. 
		This directional behavior is enforced by a substantial phase mismatch when signal and pump propagate in opposite directions. 
		However, in practical applications, unavoidable impedance mismatches when integrating TWPAs into amplification chains can result in small reflections, leading to spurious backward amplification. More precisely, considering a small non-zero reflection coefficient at both ports of the device, the effective reflection at the input port increases proportionally to the gain of the device when it is pumped \cite{ranzani_kinetic_2018}. 
		Since this backward-traveling mode shares the same frequency as the signal, conventional filtering methods are ineffective in preventing it from reaching the device under measurement, necessitating the use of isolators. 
		Consequently, the effectiveness of a truly directional amplifier can be evaluated by assessing the reflection at the input of the device at the signal frequency.
		\\

        \begin{figure}[htb]
			\centering
			\includegraphics[width=.6\textwidth]{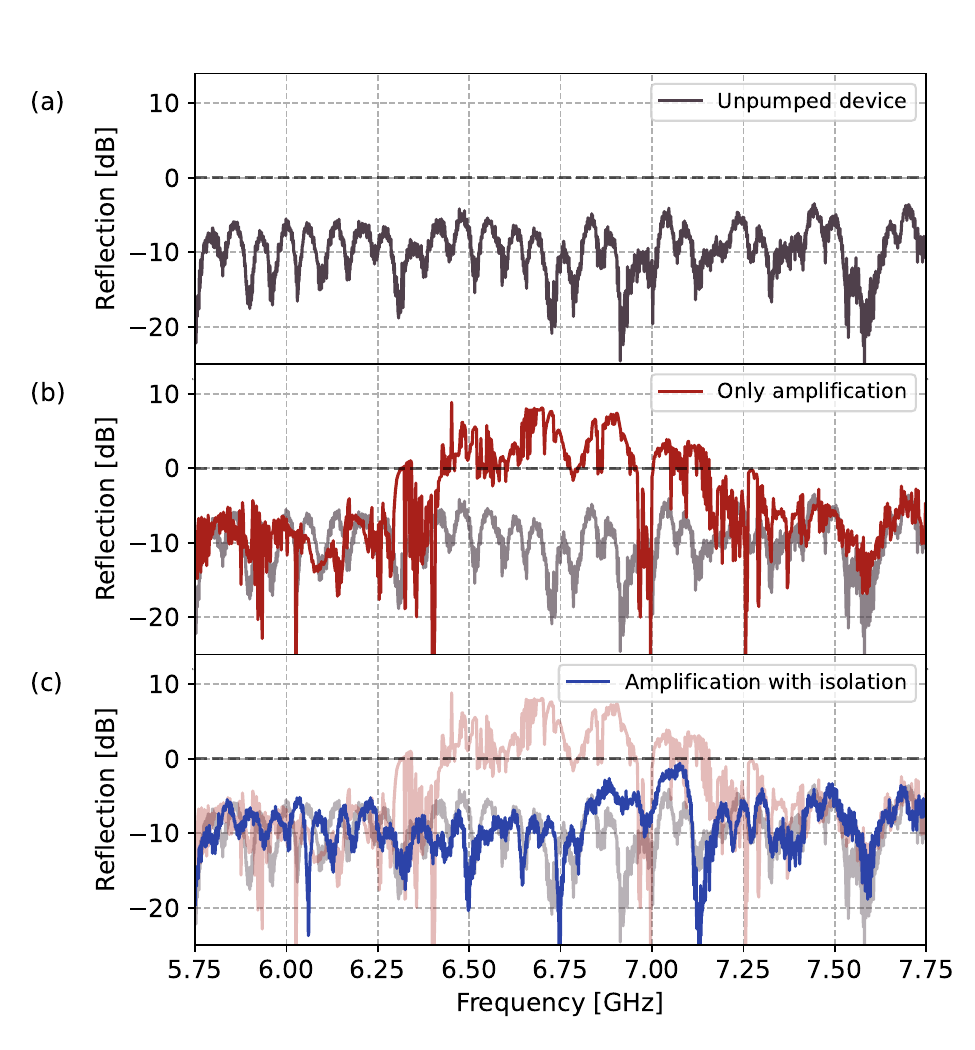}
            \caption{\textbf{Efficacy of the isolation.}
            (a) Reflection at the input port of the cold, unpumped TWPAI for reference.
            (b) Reflection at the input port of the amplifier when operated without built-in isolation, replicating the gain profile depicted in Fig \ref{fig:gain} (a).
            A significant increase is observed, primarily due to the signal reflected back after amplification caused by imperfect impedance matching.
            (c) Reflection at the input port of the TWPAI when operated in double-pumped mode with built-in isolation. 
            The reflection levels reduce significantly, approaching those of the cold reference, as backward traveling modes are upconverted and eliminated from the amplification band.
            The built-in isolation effectively mitigates the adverse effects of impedance mismatches and signal reflections, ensuring a cleaner amplification process and reducing the need for external isolators. 
            }
            \label{fig:efficacy}
		\end{figure}

		\noindent The measurement of this gain-enhanced backward reflection of the TWPAI is depicted in Fig. \ref{fig:efficacy}.
		The measurement setup was calibrated using an open reference during a separate cool-down (detailed in methods). 
		Panel (a) illustrates the reflection at the input port when the device is unpumped. It effectively serves as a cold-matched load and establishes the reference reflection for the measurement setup's microwave environment. 
		In panel (b), the reflection is shown when the amplifier operates as a standard traveling wave parametric amplifier without any backward isolation, with a gain identical to that of the TWPAI (amplifier with reverse isolation as shown in Fig. \ref{fig:gain} (a)). 
		At peak gain (red curve), an increase in reflection, corresponding to the gain minus the losses, reaching up to $+8$~dB, is observed, which can be highly detrimental to the device under measurement. 
		Finally, panel (c) illustrates (blue curve) the device operating as a TWPAI, i.e., a traveling wave parametric amplifier with in-built isolation. 
		Here, we observe a substantial reduction in the backward reflection down to the reference value of the cold-matched load. 
		This compelling evidence highlights the efficacy of the device in inhibiting the backward emission of the amplified signal, potentially eliminating the need for an isolator between the device under measurement and the first amplifier.

    \section{Conclusions}

    We have experimentally demonstrated a Josephson junction-based traveling wave parametric amplifier isolator, showcasing remarkable reverse isolation exceeding forward gain across an in-situ tunable bandwidth spanning over 500 MHz. 
	We leveraged the four-wave-mixing process for amplification and the three-wave-mixing process for backward isolation with a novel phase matching mechanism, dynamic phase matching, utilizing the presence of higher harmonics and upconverted modes.
    We demonstrate that incorporating additional parametric processes for isolation along with amplification does not compromise the system noise temperature improvement compared to a standard TWPA. 
    However, the 1~dB compression point for amplification is slightly reduced relative to a conventional four-wave-mixing TWPA utilizing Josephson nonlinearity due to the presence of a second high-power pump tone for isolation.\\

	\noindent Our investigation into the effectiveness of directional amplification in TWPAs focused on assessing the reflection input port of our device when active, serving as a key metric for evaluating the device performances. 
	We experimentally observed a substantial reduction of up to 20 dB in the backward reflection with the TWPAI configuration compared to standard TWPA operation without backward isolation. 
	This reduction opens the exciting possibility of removing the bulky magnetic isolators used between the DUTs and the first amplifier, thus paving the way towards compact, extensively multiplexed setups for measurement of quantum devices. 
    In a measurement setup, integration of TWPAIs can be facilitated using diplexers, which also separate both the pumps from the measurement circuit (detailed in supplementary information). 
    These diplexers can be integrated on-chip \cite{babenko_diplexer} with the amplifier, further adding to the drastic reduction in measurement footprint and improvement of measurement efficiency.
	The integration of the TWPAI with other quantum devices, such as qubits and superconducting circuits, offers exciting possibilities for developing fully integrated quantum systems. 
    The ability to provide both amplification and isolation in situ could simplify the design and operation of complex quantum circuits, potentially leading to new functionalities and improved overall performance.
    Moreover, the reduction in microwave components preceding the first amplifier holds promise for enhancing measurement efficiency and fidelity owing to diminished losses. \\

    \newpage
	\section{Methods}
        \subsection{The unit cell}

        The current-phase relation of a SNAIL with three large and one small junction in the loop can be expressed as a Taylor series expansion near the equilibrium flux $\phi^*$,

    \begin{equation}
    \label{eq:Taylor_SNAIL}
        \frac{I(\phi^* + \phi)}{I_\text{0}}  \approx \tilde \alpha \phi - \tilde \beta \phi^2 -\tilde \gamma \phi^3
    \end{equation}

    \noindent where, 

    \begin{equation}
    \label{eq:alpha}
        \tilde \alpha =  r \cos(\phi^*) + \cos \left( \frac{\phi^* - \phi_\text{ext}   }{3} \right)
    \end{equation}

    \begin{equation}
        \tilde \beta =  \frac{1}{2} \left[ r \sin(\phi^*) + \frac{1}{9} \sin \left( \frac{\phi^* - \phi_\text{ext}   }{3} \right)\right] 
    \end{equation}

        \begin{equation}
        \tilde \gamma =  \frac{1}{6} \left[ r \cos(\phi^*) + \frac{1}{27} \cos \left( \frac{\phi^* - \phi_\text{ext}   }{3} \right)\right] 
    \end{equation}

    \noindent The inductance per unit cell can be approximated by keeping terms up to first order in Eq \ref{eq:Taylor_SNAIL},

    \begin{equation} \label{inductance_unic_cell}
        L = \frac{\Phi_\text{0}}{2 \pi I_\text{0} \tilde \alpha}.
    \end{equation}

    \noindent The nonlinear coefficients in the main text are related to the expansion as,
    
    \begin{equation}
        \hbar  g_\text{3} = \frac{\tilde \beta}{3 \tilde \alpha } \sqrt{E_\text{c} \hbar \omega_\text{0}} \hspace{1cm}
        \hbar  g_\text{4} = \frac{\tilde \gamma}{2 \tilde \alpha } E_\text{c},
    \end{equation}

    \noindent where $\omega_\text{0}=1/\sqrt{LC_\text{g}}$ is the characteristic frequency of the transmission line, and $E_C=e^2/2C_\text{g}$ is the charging energy.
    The equation of motion in a transmission line constructed with SNAILs can be expressed as \cite{zorin_josephson_2016}: 
    
    \begin{equation}
        \frac{\partial^2 \phi}{\partial x^2} - \frac{1}{\omega_\text{0}^ 2} \frac{\partial^2 \phi}{\partial t^2} + \frac{1}{\omega_\text{J}^2} \frac{\partial^4 \phi}{\partial x^2 \partial t^2} 
        + 6 g_\text{3} \sqrt{\frac{R_\text{Q}}{\pi \omega_\text{0}^2 Z}} \frac{\partial }{\partial x} \left[ {\left(\frac{\partial \phi }{\partial x} \right) }^2 \right] 
        - 8 g_\text{4} \frac{R_\text{Q}}{\pi \omega_\text{0} Z} \frac{\partial }{\partial x} \left[ {\left(\frac{\partial \phi }{\partial x} \right) }^3 \right] = 0,
        \label{EQ_mot_SNAIL-a}
			\end{equation}

    \noindent where $\omega_\text{J} = 1/\sqrt{LC_\text{J}}$ represents the plasma frequency of the transmission line and $C_\text{J}$ is the Josephson capacitance.
    The first three terms of the equation describe the dispersion of the traveling mode in the medium. The fourth term describes second-order nonlinearity resulting in three-wave-mixing processes, and the last describes third-order nonlinearity resulting in four-wave-mixing processes.

    \subsection{Phase matching the amplification process with the dynamic phase effects}

        The linear phase mismatch between signal, idler and pump fields for four-wave-mixing amplification can be expressed as,

        \begin{equation}
            \Delta k_{\text{dispersion}} = 2k_\text{ap} - k_\text{s} - k_\text{i},
            \label{lin_delta_k}
        \end{equation}

        \noindent where $k_\text{s}$, $k_\text{i}$, and $k_\text{ap}$ are wavevectors defined by the linear dispersion in the equation of motion ($g_\text{3}=g_\text{4}=0$ in Eq. \ref{EQ_mot_SNAIL-a}), 

        \begin{equation}
            k_\text{m}(\omega) = \frac{\omega_\text{m}}{\omega_\text{0} \sqrt{1 - \omega_\text{m}^2/\omega_\text{J}^2}}.
        \end{equation}

        \noindent where $\omega_\text{m}$ is the frequency of the mode.
        The Kerr phase mismatch for the process stems from the presence of self-Kerr and cross-Kerr effect and can be quantified as \cite{ranadive_kerr_2022},

        \begin{equation}
            \Delta k_{\text{Kerr}} = 2\eta_\text{ap} - \eta_\text{s} - \eta_\text{i},
            \label{eq:etas2}
        \end{equation}

        \noindent with,

        \begin{equation}
            \begin{gathered}
                \eta_\text{{s,i}} = \frac{6\gamma}{8\widetilde{\omega}_\text{{s,i}}} k_\text{ap}^2 k_\text{{s,i}} |A_\text{ap}|^2 , \hspace{0.5cm}
                \eta_\text{ap} = \frac{3\gamma}{8\widetilde{\omega}_\text{ap}} k_\text{ap}^3 |A_\text{ap}|^2 ,
            \end{gathered}
            \label{eq:etas}
        \end{equation}

        \noindent where $|A_\text{ap}|$ is the pump amplitude, $\widetilde{\omega}_\text{m}=\left(1-\omega_\text{m}^2/\omega_\text{J}^2\right)$ and $\gamma = 8 g_\text{4} \frac{R_\text{Q}}{\pi \omega_\text{0} Z}$.\\

        \noindent The presence of second-order nonlinearity ($g_\text{3}$), results in a Kerr-like effect on the evolution of the phase of the waves traveling in the nonlinear medium. 
        A simple expression can be approximated for this effect by solving the Eq. \ref{EQ_mot_SNAIL-a} in absence of third order nonlinearity.
        The equation for second harmonic generation (SHG) can be approximated as,

        \begin{equation}
            \frac{\partial A_\text{2ap}}{\partial x} = \frac{\beta}{4\widetilde{\omega}_\text{2ap}} k_\text{ap}^2 A_\text{ap}^2 e^{i\Delta k_\text{2ap} x}
        \end{equation}

        \noindent where $\omega_\text{2ap}=2\omega_\text{ap}$ is the frequency of the second harmonic, $\beta = 6 g_\text{3} \sqrt{\frac{R_\text{Q}}{\pi \omega_\text{0}^2 Z}}$ and $ \Delta k_\text{2ap} = 2 k_\text{ap} - k_\text{2ap}$; this equation can be solved trivially with stiff pump approximation (constant $|A_\text{ap}|^2$) and low dynamic phase modulation ($\Delta k_\text{2ap} \gg \eta_\text{dynamic}$),
        
        \begin{equation}
            A_\text{2ap}(x) = \frac{-i\beta}{4 \Delta k_\text{2ap} \widetilde{\omega}_\text{2ap}} k_\text{ap}^2 A_\text{ap}^2 e^{i\Delta k_\text{2ap} x}
        \end{equation}

        \noindent The evolution of the pump can then be approximated as,

        \begin{equation}
            \frac{\partial A_\text{ap}}{\partial x} = \frac{i\beta^2}{8 \Delta k_\text{2ap} \widetilde{\omega}_\text{ap} \widetilde{\omega}_\text{2ap}} k_\text{2ap} k_\text{ap}^3 A_\text{ap}^2 A_\text{ap}^*  
        \end{equation}

        \noindent and a Kerr-like term can be extracted as,

        \begin{equation}
            \eta_\text{ap,dyn} = \frac{\beta^2}{8 \Delta k_\text{2ap} \widetilde{\omega}_\text{ap} \widetilde{\omega}_\text{2ap}} k_\text{2ap} k_\text{ap}^3 |A_\text{ap}|^2
            \label{dyn_coef_pump}
        \end{equation}

        \noindent Similarly, the equation for the upconverted signal mode can be approximated as,
        
        \begin{equation}
            \frac{\partial A_\text{s+p}}{\partial x} 
                = \frac{\beta}{2\widetilde{\omega}_\text{s+ap} } k_\text{ap} A_\text{ap} k_\text{s} A_\text{s} e^{i\Delta k_\text{s+ap} x}
        \end{equation}

        \noindent where $\widetilde \omega_\text{s+ap}=\omega_\text{s}+\omega_\text{ap}$ is the frequency of the upconverted signal mode and $\Delta k_\text{s+ap} = k_\text{ap} + k_\text{s} - k_\text{s+ap}$; this equation can be solved trivially with constant signal approximation ($|A_\text{s}|^2$ constant as the process is not phase matched) and low dynamic phase modulation ($\Delta k_\text{s+ap} \gg \eta_\text{dynamic}$),

        \begin{equation}
            A_\text{s+ap}(x) = \frac{-i\beta}{2 \Delta k_\text{s+ap} \widetilde{\omega}_\text{s+ap}} k_\text{ap} A_\text{ap} k_\text{s} A_\text{s} e^{i\Delta k_\text{s+ap} x}
        \end{equation}

        \noindent The evolution of the signal can then be approximated as,

        \begin{equation}
            \frac{\partial A_\text{s}}{\partial x} = \frac{i\beta^2}{4 \widetilde{\omega}_\text{s} \widetilde{\omega}_\text{s+ap} \Delta k_\text{s+ap}} k_\text{ap}^2 |A_\text{ap}|^2 k_\text{s+ap} k_\text{s} A_\text{s}
        \end{equation}

        \noindent and a Kerr-like term can be extracted as,

        \begin{equation}
            \eta_\text{s,dyn} = \frac{\beta^2}{4 \widetilde{\omega}_\text{s} \widetilde{\omega}_\text{s+ap} \Delta k_\text{s+ap}} k_\text{ap}^2 |A_\text{ap}|^2 k_\text{s+ap} k_\text{s}
            \label{dyn_coef_sig}
        \end{equation}

        \noindent Similarly for the idler mode,

        \begin{equation}
            \eta_\text{i,dyn} = \frac{\beta^2}{4 \widetilde{\omega}_\text{i} \widetilde{\omega}_\text{i+ap} \Delta k_\text{i+ap}} k_\text{ap}^2 |A_\text{ap}|^2 k_\text{i+ap} k_\text{i}
            \label{dyn_coef_idler}
        \end{equation}

        \noindent A correction to the above approximations can be made by considering the power drain of the pump mode to the second harmonic, and appropriately modifying the pump power ($|A_\text{ap}|^2$). \\

        \noindent The dynamic phase mismatch for the process can be quantified as,

        \begin{equation}
            \Delta k_\text{dynamic} = 2 \eta_\text{ap,dyn} - \eta_\text{s,dyn} - \eta_\text{i,dyn}
            \label{dyn_delta_k}
        \end{equation}

        \noindent And the total phase mismatch for the wave mixing process can be computed as the aggregate of the three components, 

        \begin{equation}
            \Delta k = \Delta k_\text{dispersion} + \Delta k_\text{Kerr} + \Delta k_\text{dynamic}.
        \end{equation}
        \\

        \noindent The circuit parameters, $C_\text{g}=250$~fF, $C_\text{J}=35$~fF, and loss tangent $\tan(\delta)=2.9\times 10^{-3}$ used for computing the curves depicted in panel (d) of Fig. \ref{fig:gain} using above expressions, were obtained from the design and linear characterization of the device.
        The remaining parameters, the three-wave ($g_\text{3}$) and four-wave ($g_\text{4}$) mixing nonlinear coefficients, were obtained using best-fit for the gain curve as detailed in the next section.
        The effect of pump attenuation due to presence of dielectric losses was modeled as a position independent amplitude reduction as, $|A_\text{p}| = A_\text{{p0}} \exp(-k_\text{p} \tan(\delta) l/4)$ \cite{macklin_nearquantum-limited_2015}, where $l$ is length of the TWPAI.
        
	\subsection{Coupled mode equations for the dynamically phase matched gain}

     The amplification process employed in the TWPAI is driven by a four-wave-mixing interaction. 
     In this process, two pump photons combine to produce one signal photon and one idler photon ($2\omega_\text{ap} = \omega_\text{s} + \omega_\text{i}$). 
     To capture this process, it is essential to account for at least three initial modes: the signal ($\omega_\text{s}$), the pump ($\omega_\text{ap}$), and the idler ($\omega_\text{i}$).
     However, due to the involvement of second-order nonlinearity and its dynamic impact on phase evolution within the amplification process, corresponding upconversion phenomena must also be considered to accurately depict the gain behavior. 
     Consequently, the equation of motion within the nonlinear medium (Eq. \ref{EQ_mot_SNAIL-a}) needs to be solved with a six-mode ansatz, the signal, idler, pump, second-harmonic of the pump ($\omega_\text{2ap} = 2\omega_\text{ap}$), upconverted signal ($\omega_\text{s+ap} = \omega_\text{s} + \omega_\text{ap}$), and upconverted idler ($\omega_\text{i+ap} = \omega_\text{i} + \omega_\text{ap}$).
     This results in a system of six coupled mode equations, 
     
    \begin{multline}
        \frac{\partial A_\text{s}}{\partial x} =  \frac{6 \gamma i k_\text{s}}{8 \widetilde \omega_\text{s}}  \Biggl[ k_\text{ap}^2 |A_\text{ap}|^2 +  k_\text{2ap}^2 |A_\text{2ap}|^2 \Biggr] A_\text{s} - \frac{3 \gamma  i }{8 \widetilde \omega_\text{s} k_\text{s}} (-2k_\text{ap} + k_\text{i})k_\text{ap}^2 k_\text{i}  A_\text{ap}^2 A_\text{i}^* e^{i(2k_\text{ap} - k_\text{i} -k_\text{s})x} \\
        + \beta \frac{k_\text{ap} - k_\text{s+ap} }{2 k_\text{s}\widetilde \omega_\text{s}} k_\text{ap} k_\text{s+ap} A_\text{s+ap} A_\text{ap}^* e^{i(k_\text{s+ap} - k_\text{ap} - k_\text{s})x} - k_\text{s}^{''}A_\text{s}
    \end{multline}
    
    \begin{equation}
         \frac{\partial A_\text{ap}}{\partial x} = \beta  \frac{(k_\text{ap} - k_\text{2ap})}{2\widetilde \omega_\text{ap} }  k_\text{2ap}  A_\text{ap}^* A_\text{2ap} e^{i(k_\text{2ap} - 2k_\text{ap})x} + \frac{3\gamma  i k_\text{ap}}{8\widetilde \omega_\text{ap} } \Biggl[ k_\text{ap}^2 |A_\text{ap}|^2   + 2  k_\text{2ap}^2  |A_\text{2ap}|^2  \Biggr] A_\text{ap} - k_\text{ap}^{''}A_\text{ap}
    \end{equation}
    
    \begin{multline}
        \frac{\partial A_\text{i}}{\partial x} =  \frac{6 \gamma i k_\text{i}}{8 \widetilde \omega_\text{i}}  \Biggl[ k_\text{ap}^2 |A_\text{ap}|^2 +  k_\text{2ap}^2 |A_\text{2ap}|^2 \Biggr] A_\text{i} - \frac{3 \gamma i  }{8 \widetilde \omega_\text{i} k_\text{i}} (-2 k_\text{ap} + k_\text{s})k_\text{ap}^2 k_\text{s} A_\text{ap}^2 A_\text{s}^*   e^{i(2k_\text{ap} - k_\text{i} -k_\text{s})x}\\
        +  \beta \frac{k_\text{ap} - k_\text{i+ap} }{\widetilde \omega_\text{i}} k_\text{i+ap} k_\text{ap} A_\text{i+ap} A_\text{ap}^* e^{i(k_\text{i+ap} - k_\text{ap} - k_\text{i})x} - k_\text{i}^{''}A_\text{i}
    \end{multline}

    \begin{equation}
         \frac{\partial A_\text{2ap}}{\partial x} =  \beta \frac{k_\text{ap}^3}{2 k_\text{2ap} \widetilde \omega_\text{2ap} }   A_\text{ap}^2 e^{i(2k_\text{ap} - k_\text{2ap})x} + \frac{3\gamma  i k_\text{2ap}}{8\widetilde \omega_\text{2ap} } \Biggl[2 k_\text{ap}^2 |A_\text{ap}|^2   +   k_\text{2ap}^2  |A_\text{2ap}|^2  \Biggr] A_\text{2ap} -  k_\text{2ap}^{''}A_\text{2ap}
    \end{equation}
    
    \begin{equation}
        \frac{\partial A_\text{s+ap} }{\partial x} =  \beta \frac{k_\text{s} + k_\text{ap} }{2 k_\text{s+ap} \widetilde \omega_\text{s+ap} } k_\text{ap} k_\text{s} A_\text{s} A_\text{ap} e^{i(k_\text{s} + k_\text{ap} - k_\text{s+ap})x} +  \frac{6\gamma  i k_\text{s+ap}}{8\widetilde \omega_\text{s+ap} } \Biggl[ k_\text{ap}^2 |A_\text{ap}|^2   +   k_\text{2ap}^2  |A_\text{2ap}|^2  \Biggr] A_\text{s+ap} - k_\text{s+ap}^{''}A_\text{s+ap}
    \end{equation}
    
    \begin{equation}
        \frac{\partial A_\text{i+ap} }{\partial x}  =  \beta \frac{k_\text{ap} + k_\text{i}}{2 k_\text{i+ap}\widetilde \omega_\text{i+ap}} k_\text{ap} k_\text{i} A_\text{i} A_\text{ap} e^{i(k_\text{i} + k_\text{ap} - k_\text{i+ap})x} +  \frac{6\gamma  i k_\text{i+ap}}{8\widetilde \omega_\text{i+ap} } \Biggl[ k_\text{ap}^2 |A_\text{ap}|^2   +   k_\text{2ap}^2  |A_\text{2ap}|^2  \Biggr] A_\text{i+ap} - k_\text{i+ap}^{''}A_\text{i+ap}
    \end{equation}

    \noindent where  $k^{''}$ is the imaginary part of the wavevector, $ k^{''} = k \tan(\delta)/2$.\\

    \noindent The circuit parameters, $C_\text{g}=250$~fF, $C_\text{J}=35$~fF, and loss tangent $\tan(\delta)=2.9\times 10^{-3}$ used for simulating the gain depicted in panel (b) of Fig. \ref{fig:gain} by solving the above system of equations, were obtained from the design and linear characterization of the device.
    The remaining parameters, the three-wave ($\beta$) and four-wave ($\gamma$) mixing nonlinear coefficients, were obtained to best-fit the gain curve.
    The fitting process becomes essential because the simplified expressions for unit cells do not precisely encapsulate the flux dependence of the underlying nonlinearities in the flipped SNAIL metamaterial. 
    While the full circuit simulator WRspice adeptly captures this behavior, CME simulation is included to provide a more intuitive understanding of the underlying mechanisms driving the amplification process.

	\subsection{WRspice simulation}

    \begin{figure}[htb]
        \centering
        \includegraphics[scale = 0.8]{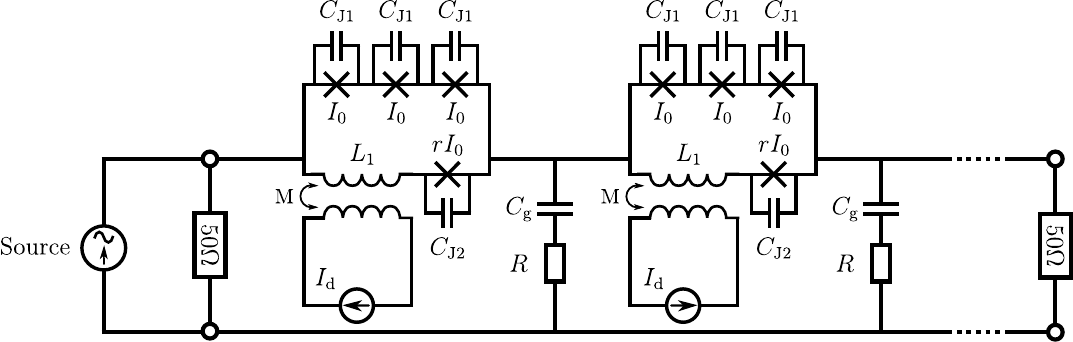}
        \caption{\textbf{The circuit used for WRspice simulations.}
        The losses were modeled as equivalent series resistors with the ground capacitance. 
        Mutual inductances were used to introduce the flux effect from an external magnetic field across the SNAIL.
        The orientation of the SNAIL relative to the external magnetic flux was modeled by the direction of the direct current in the inductors.}
        \label{WRspice_circuit}
    \end{figure}

   The circuit used in WRspice simulation is depicted in Fig \ref{WRspice_circuit}.
   The dielectric losses in the device (loss tangent) were introduced in the circuit as equivalent series resistors.
   The circuit parameters, $r=0.066$, $C_\text{g}=250$~fF, $C_\text{J1}=76.5$~fF, $C_\text{J2}=5.36$~fF, and loss tangent $\tan(\delta)=2.9\times 10^{-3}$ used for the simulation were obtained from the design and linear characterization of the device, with the exception of $I_\text{0}=2.2 \: \text{$\mu$A}$, which needed to be set at about 20$\%$ above the fitted value to reproduce experimentally the observed behavior discussed in Fig \ref{fig:phase_matching} and Fig. \ref{fig:isolation} of the main text.
   The inherent uncertainties in the linear measurements and the limitation of the simplified SNAIL model could have contributed to this deviation, and a more comprehensive modeling and analysis is beyond the scope of this article.

	\subsection{The measurement setup and its calibration}
  
    \begin{figure}[htb]
        \centering
        \includegraphics[width=.85\textwidth]{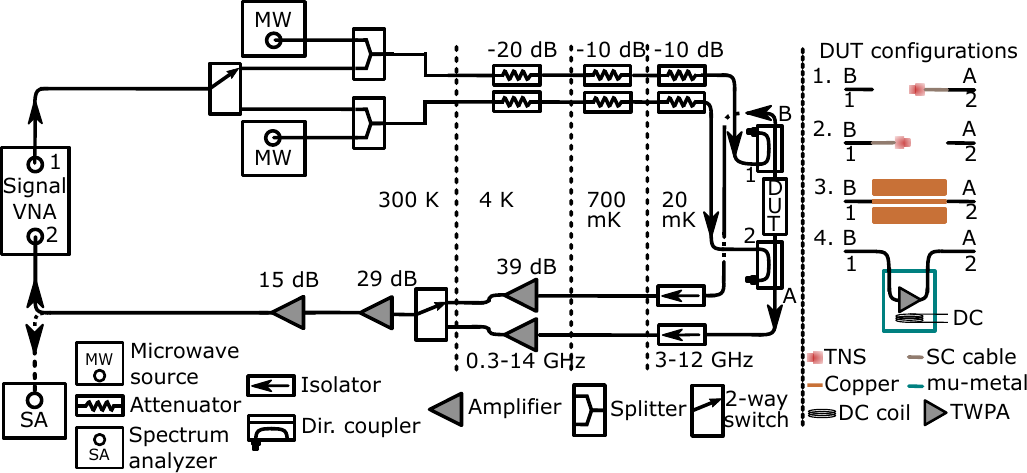}
        \caption{\textbf{The experimental setup and measurement configurations.}
        The experimental setup allowing for forward and backward characterization is shown on the left along with the different configurations used at the base (20~mK) of the cryostat for the calibrations and the measurements on the right. }
        \label{exp_setup}
    \end{figure}

    For a comprehensive characterization of the TWPAI, it was essential to employ a setup capable of measuring the full S-matrix (forward and backward transmission along with reflection) of the device while simultaneously injecting microwaves in both forward and reverse directions. 
    Achieving this capability required the integration of directional couplers before and after the device, as illustrated in Fig. \ref{exp_setup}. 
    Consequently, the setup comprised two input lines (labeled 1 and 2 in Fig. \ref{exp_setup}) and two output lines (labeled A and B in Fig. \ref{exp_setup}). 
    Three separate cooldowns were conducted to calibrate the measurement setup and establish a reference, each with distinct configurations as depicted in the right panel of Fig. \ref{exp_setup}. \\
    
    \noindent In the first configuration, the directional coupler connected to input 1 and output B is left open, while the directional coupler connected to input 2 and output A is connected to a thermally controlled noise source. 
    This setup allows for the calibration of gain and noise of output line A while providing the necessary data for reflection calibration at port 1.
    In the second measurement run, the configuration is reversed: the directional coupler connected to input 1 and output B is now connected to the noise source, and the coupler connected to input 2 and output A is left open. 
    This reversal enables the calibration of gain and noise of output line B and gathers the data required for reflection calibration at port 2.
    By combining the data from these two distinct measurement runs, the reference plane for the power measurements can be set directly at the TWPAI ports.\\
    
    \noindent In the third configuration, a sample box containing a coplanar waveguide (CPW) printed circuit board (PCB), positioned in place of the TWPAI, is connected between the two directional couplers. 
    This configuration facilitates the measurement of a phase reference, which is essential for the linear characterization of the TWPAI. 
    The phase reference provides a baseline against which the dispersion in the TWPAI can be compared.
    The methodology for the linear characterization is detailed in \cite{ranadive_nonlinear_2023}.\\
    
    \noindent The output lines' noise and gain were calibrated using a broadband thermal noise source (TNS) with adjustable temperatures ranging from 40~mK to 1~K. 
    The TNS comprises an impedance-matched resistive termination mounted on a copper platform with controllable temperature and connected to the output line with a superconducting cable, ensuring minimal loss and thermal conduction \cite{ranadive_nonlinear_2023}.
    The power spectral density of the noise emitted by the TNS was measured across different temperatures using a spectrum analyzer to characterize the noise performance. 
    The collected data was then fitted to the noise model,

    \begin{equation}
        P(\omega_\text{s}) = \{ 
            N_{\text{source}}(\omega,T_{\text{source}}) \, + N_{\text{out}}(\omega)
        \} \,  G_{\text{out}}(\omega) B_\text{w}  \, , 
        \label{noise_model_ref}
    \end{equation}

    \noindent where, $G_{\text{out}}(\omega)$ and  $N_{\text{out}}(\omega)$ are the fit parameters corresponding to the output line, $B_{\text{w}}$ is the measurement bandwidth, and $N_\text{source}(\omega,T_\text{source})$ is the noise emitted by the thermal noise noise at frequency $\omega$, when heated to temperature $T_\text{source}$:

    \begin{equation}
        N_\text{source}(\omega,T_\text{source}) = \frac{\hbar \omega}{2} \coth \left(  \frac{\hbar \omega}{2 k_B T_\text{source}} \right).
    \end{equation}

    \subsection{Losses in the device}

        \begin{figure}[htb]
        \centering
        \includegraphics[width = 0.7\textwidth]{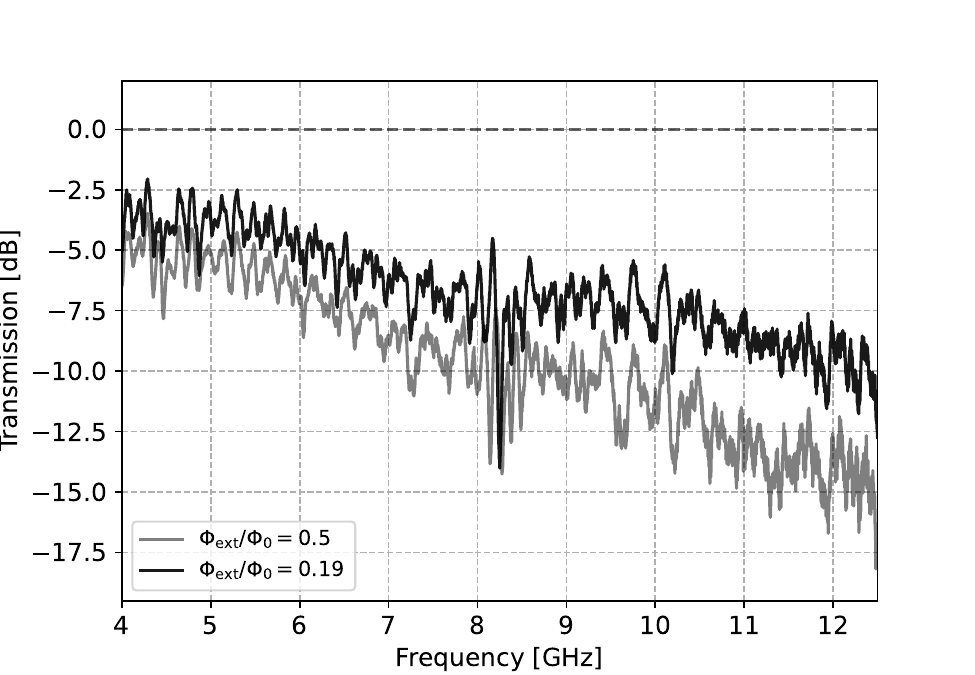}
        \caption{Low power ($\sim 1$ photon) dissipation in the TWPAI at two selected flux values, $\Phi_{\text{ext}} =  0.19$~$ \Phi_\text{0}$ and $\Phi_{\text{ext}} = 0.5$~$\Phi_\text{0}$}
        \label{loss}
    \end{figure}

    Figure \ref{loss} shows transmission losses of the device for two selected flux values, $\Phi_{\text{ext}} =  0.19$~$ \Phi_\text{0}$ and $\Phi_{\text{ext}} = 0.5$~$\Phi_\text{0}$ in low power ($\sim 1$ photon) regime. 
    These frequency-dependent dielectric losses can be modeled with loss tangent, i.e., the ratio of the imaginary part to the real part of the metamaterial's complex permittivity as,

    \begin{equation}
        |A(x)| = |A(0)| e^{-k(\omega) \tan(\delta)x/2}
    \end{equation}
    
    \noindent where $k(\omega)$ is the wave-vector and $\tan(\delta)$ is the loss tangent. 
    The loss tangent can be determined by fitting the observed losses to the model \cite{planat_fabrication_2019, ranadive_kerr_2022}. 
    Ideally, the loss tangent should be independent of the applied magnetic flux. 
    However, in practice, the fit parameter shows a variation of approximately 20 percent around the central value of $2.9 \times 10^{-3}$. 

	\subsection{Tunability}

\begin{figure}[htb]
        \centering
        \includegraphics[width=.8\textwidth]{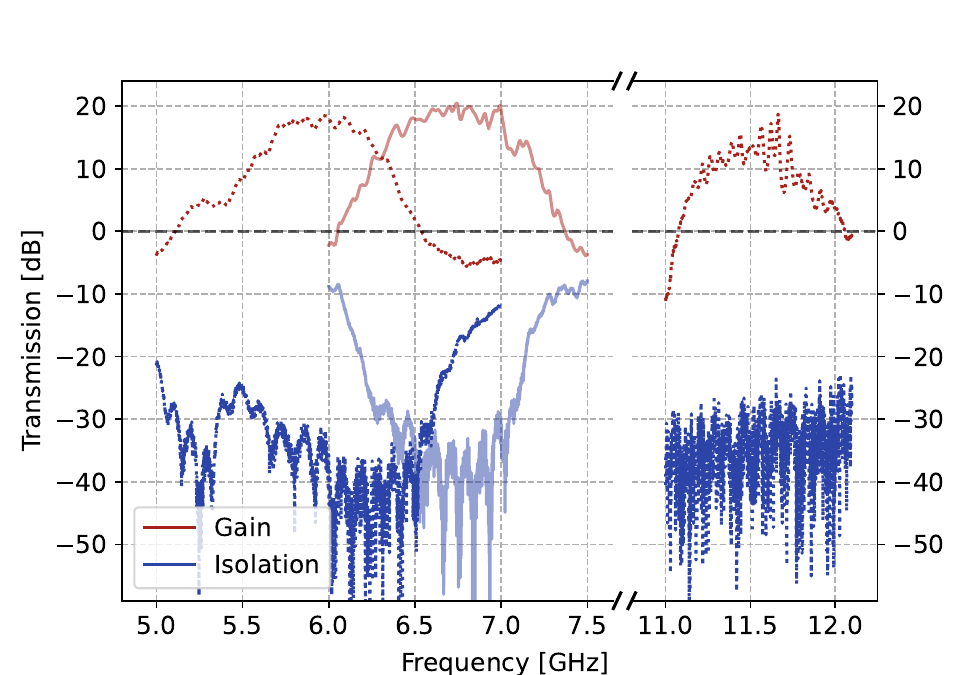}
        \caption{\textbf{In situ gain and isolation band tunability.}
        Various gain and isolation profiles were measured by simply adjusting the corresponding pump frequencies. 
        The red curves represent the gain, while the blue curves indicate the corresponding isolation. 
        The forward (amplification) pump frequencies used were 9~GHz, 9.5~GHz, and 9.15~GHz from left to right. 
        Similarly, the backward (isolation) pump frequencies were 13.71~GHz, 13.85~GHz, and 7.57~GHz in the same order. 
        The external flux applied were $\Phi_{\text{ext}} =$ 0.26, 0.19, and 0.35~$\Phi_\text{0}$, respectively.
        }
        \label{tunability}
    \end{figure}

   A key advantage of the TWPAI is its capability for in situ tuning of amplification and isolation bands over a range of gigahertz.
    Figure \ref{tunability} presents the gain profiles and corresponding isolation profiles for three different pairs of gain and isolation pump frequencies.
    These profiles can be adjusted almost continuously by varying the pump frequencies.
    The transmission curves are referenced to a lossless line.
    The device achieves a gain of up to 20~dB for frequencies up to 8~GHz; however, this gain diminishes at higher frequencies due to increased dielectric losses.

    \newpage

	\bibliographystyle{apsrev4-2}
	
    \bibliography{lib1_v2,lib2,lib3}

	\section{Acknowledgements}

        This project has received funding from European Union's Horizon Europe 2021-2027 project TruePA (grant agreement number 101080152) and from the French ANR-22-PETQ-0003 grant under the 'France 2030' plan.
        B.F. acknowledges the QMIC project under the program DOS0195438/00.
        We would like to acknowledge M. Esposito for her significant assistance with the sample fabrication and for enlightening discussions.
        The samples were fabricated at the clean room facility \textit{Nanofab} of Institut Néel in Grenoble. 
        We thank the clean room staff and L. Cagnon for their assistance with the device fabrication.
        We express our gratitude to J. Jarreau, L. Del Rey, D. Dufeu, F. Balestro, and W. Wernsdorfer for their support with the experimental equipment. 
        We are also grateful to the superconducting quantum circuits group members at Néel Institute for helpful discussions.
        We thank R. Albert, O. Buisson, A. Coissard, M. Esposito and Q. Ficheux for their critical reading and constructive feedback on the manuscript.
        
	\section{Author contributions}

        A.R., B.F., and N.R. conceptualized the experiment, A.R. fabricated the device and measured in the cryogenic setup with the help of G.L.G., B.F., G.C., and G.B., while L.P., E.B., E.E. and N.R. provided support with the measurement setup.
        A.R. and B.F. analyzed the measurement data with the help of G.L.G., G.C. and N.R.
        A.R. and B.F. performed the simulations with inputs from G.L.G., S.B., A.M., and N.R.
        A.R., B.F., G.L.G., and N.R. drafted the article with contributions from all the authors.
        
	\section{Competing interests}
         N.R. and L.P are founders and shareholders of Silent Waves. 
         A.R., B.F., and G.B. are equity and/or options holders in Silent Waves.
         The authors associated with Silent Waves have financial interests in the company.

\end{document}


\title{Supplementary information: A Traveling Wave Parametric Amplifier Isolator}  \author{Arpit Ranadive} 
    \affiliation{Univ. Grenoble Alpes, CNRS, Grenoble INP, Institut N\'eel, 38000 Grenoble, France}
    \author{Bekim Fazliji} 
    \affiliation{Univ. Grenoble Alpes, CNRS, Grenoble INP, Institut N\'eel, 38000 Grenoble, France}
    \affiliation{Silent Waves, 38000 Grenoble, France}
    \author{Gwenael Le Gal} 
    \affiliation{Univ. Grenoble Alpes, CNRS, Grenoble INP, Institut N\'eel, 38000 Grenoble, France}
    \author{Giulio Cappelli} 
    \affiliation{Univ. Grenoble Alpes, CNRS, Grenoble INP, Institut N\'eel, 38000 Grenoble, France}
    \author{Guilliam Butseraen} 
    \thanks{Now at Silent Waves.}
    \affiliation{Univ. Grenoble Alpes, CNRS, Grenoble INP, Institut N\'eel, 38000 Grenoble, France}
    \author{Edgar Bonet} 
    \affiliation{Univ. Grenoble Alpes, CNRS, Grenoble INP, Institut N\'eel, 38000 Grenoble, France}
    \author{Eric Eyraud} 
    \affiliation{Univ. Grenoble Alpes, CNRS, Grenoble INP, Institut N\'eel, 38000 Grenoble, France}
    \author{Sina B\"{o}hling} 
    \affiliation{Institute for Theory of Condensed Matter and Institute for Quantum Materials and Technology, Karlsruhe Institute of Technology, 76131 Karlsruhe, Germany}  
    \author{Luca Planat}
    \affiliation{Silent Waves, 38000 Grenoble, France}
    \author{A. Metelmann} 
    \affiliation{Institute for Theory of Condensed Matter and Institute for Quantum Materials and Technology, Karlsruhe Institute of Technology, 76131 Karlsruhe, Germany} 
    \affiliation{Institut de Science et d’Ingénierie Supramoléculaires (ISIS, UMR7006), University of Strasbourg and CNRS}
    \author{Nicolas Roch} 
    \thanks{Corresponding author: nicolas.roch@neel.cnrs.fr}
    \affiliation{Univ. Grenoble Alpes, CNRS, Grenoble INP, Institut N\'eel, 38000 Grenoble, France}
    \affiliation{Silent Waves, 38000 Grenoble, France}
	
	\maketitle

    \section{Experimental setup with diplexers}

    \begin{figure}[htbp]
        \centering
        \includegraphics[width=.75\textwidth]{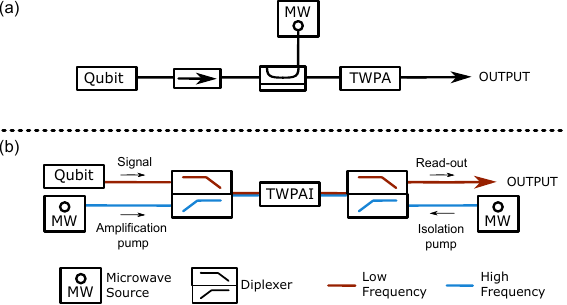}
        \caption{\textbf{Setup improvement.} (a) Standard setup for qubit measurement. (b) Proposed setup with diplexers and the TWPAI. The frequency domain is split in two: low frequency (blue) and high frequency (red)}.
        \label{fig:enter-label}
    \end{figure}

    The TWPAI offers significant advantages due to its built-in reverse isolation. 
    A primary advantage is eliminating the need for commercial isolators in standard qubit measurement setups by leveraging the TWPAI's inherent isolation capabilities. 
    Utilizing the TWPAI in the regime presented in the article requires a setup enabling double pumping. 
    However, the isolation pump, which travels backward toward the qubit, might impact the device under test. To mitigate this, a diplexer can be used to separate the backward-traveling pump.\\
    
    For example, with a qubit having a readout resonator frequency of around 7~GHz, the TWPAI can be operated, as shown in Figure 4 of the article, with the amplification and isolation pumps positioned above 9~GHz. 
    By placing two diplexers with a cutoff frequency at 8 GHz before and after the TWPAI, the frequency domain is effectively split into low ($<$ 8~GHz) and high ($>$ 8~GHz) ranges. 
    In this configuration, the two pumps reside in the high-frequency domain, while the device under test remains in the low-frequency domain. 
    This setup also separates the amplification pump from the amplification chain following the TWPAI.\\
    
    Moreover, using diplexers generally results in lower losses compared to standard isolators. 
    This loss reduction between the device under test and the TWPAI is expected to enhance qubit readout performance. 
    This discussion extends beyond qubit measurement and applies to low-noise microwave measurements in general.

	\bibliographystyle{apsrev4-2}

    \bibliography{lib1,lib2,lib3}